\documentclass[12pt]{article}
\usepackage{epsfig}
\usepackage{amsmath}

\newcommand{\eden}{\varepsilon}

\begin{document}
\begin{titlepage}

\begin{flushright}
CERN-TH/2003-014
\end{flushright}

\medskip
\begin{center}
{\large \bf On energy densities reached in\\
 heavy-ion collisions at the CERN SPS}
\end{center}
\vspace{1cm}
\begin{center}
{ J\'an Pi\v{s}\'ut${}^{a,b}$, Neva Pi\v{s}\'utov\'a${}^{b}$,
and Boris Tom\'a\v{s}ik${}^{a}$}

\vspace{1cm}

${}^{a}${\it CERN, Theory Division, CH-1211 Geneva 23, Switzerland}
\end{center}
\begin{center}
${}^{b}${\it Department of Physics, Comenius University,
 SK-84248 Bratislava, Slovakia}
\end{center}
\vspace{1cm}
\begin{center}
March 13, 2003
\end{center}
\vspace{1cm}
\abstract
{We present a few estimates of energy densities reached in
heavy-ion collisions at the CERN SPS. The estimates are based 
on data and models of proton--nucleus and nucleus--nucleus
interactions. In all of these estimates the maximum energy density
in central Pb+Pb interactions is larger than the critical energy
density $\eden_c\approx 0.7\, \mbox{GeV/fm}^{3}$ following from
lattice gauge theory computations. In estimates which we consider as
realistic the maximum energy density is about $2\eden_c$. 
In this way our analysis gives some support to claims that 
deconfined matter has been produced
at the CERN SPS. Any definite statement requires a deeper understanding
of formation times of partons and hadrons in nuclear collisions.
We also compare our results with implicit energy estimates
contained in earlier models of anomalous  $J/\psi$ suppression
in nuclear collisions.}
\end{titlepage}


\section{Introduction}
\label{intro}

There exist plenty of models and Monte Carlo generators of 
proton--proton (pp), proton--nucleus (pA) and 
nucleus--nucleus (AB)
interactions. These include models based on strings and their 
fragmentation, e.g. \cite{LUND,DPM,VENUS,QMD}, on partonic cascades
\cite{GM,Geiger}, on hadronic cascades \cite{Hum,Kahana}, 
on combined parton and hadron degrees of freedom 
\cite{COMB}, and on other
pictures of the initial state of the collision. After fixing
a few parameters, these models are able to find a reasonable 
agreement with data.

Another class of models for pA interactions is based on
successive collisions of the incident proton 
with those nucleons in A which are present within a tube in A
given by the trajectory of the proton in the nucleus and by the
non-diffractive cross-section for proton--nucleon interaction, e.g.\
\cite{Date,Wong1,Wong2,Wong3,Sumi1,Sumi2,Lexus,Satz1,Liet1, Liet2,Lich}.
The model can be naturally extended to AB interactions. There it 
is based on the Glauber picture of colliding tubes of 
nucleons, see e.g.\ \cite{Wong1,Wong2,Wong3,Sumi1,Sumi2,Lexus,Satz1}.
Models of this type proceed at every step in accordance with the
data available from pp and pA interactions. A crucial input comes 
from data on the formation time of hadrons and on the Drell--Yan process
\cite{Gale}. Such a method  keeps
the parameters under control and gains some information about the 
space-time evolution of the process. 

One of the most important quantities of interest in heavy-ion 
collisions is the highest energy density reached. This quantity is
relevant to the possible approach to the quark--gluon plasma.
In this paper we estimate the energy density reached at the CERN SPS 
by using a simple version of the Glauber-type model with 
tube-on-tube interactions.

We shall compare our estimates with recent lattice gauge 
theory results \cite{Karsch1,Karsch2}. The lattice results
have brought a new and most interesting information on the type
and parameters of the phase transition between the quark-gluon plasma
(QGP) and the hadron gas (HG). The phase transition is most likely 
of a cross-over type, with critical temperature 
$T_c\approx 173\, \mbox{MeV}$ and the rather low critical energy density of
$\eden_c=\eden(T_c)\approx 0.7 \, \mbox{GeV/fm}^{3}$.

The paper is organized as follows. In Section~\ref{model} we describe 
a simple version of the Glauber-type model of AB interactions, which
we use in our calculations of rapidity distribution of the energy  
$\Delta E/\Delta y$. We also present our estimate of the volume 
  occupied by the central rapidity region of $\Delta y\approx 1$ 
right after the tube-on-tube interaction
is finished.
The energy density is then estimated as the ratio of 
$\Delta E/\Delta y$ and $\Delta V/\Delta y$. 
More details of the model are explained in Section~\ref{comments}.
In particular we  discuss
the formation time of hadrons and its relationship to
the space-time picture of subsequent energy losses of incident nucleons
in a tube-on-tube interaction. 
In Section~\ref{results} we present our results. Comments and concluding 
remarks are deferred to the last section.


\section{The model}
\label{model}     

\begin{figure}[t]
\begin{center}
   \epsfig{file=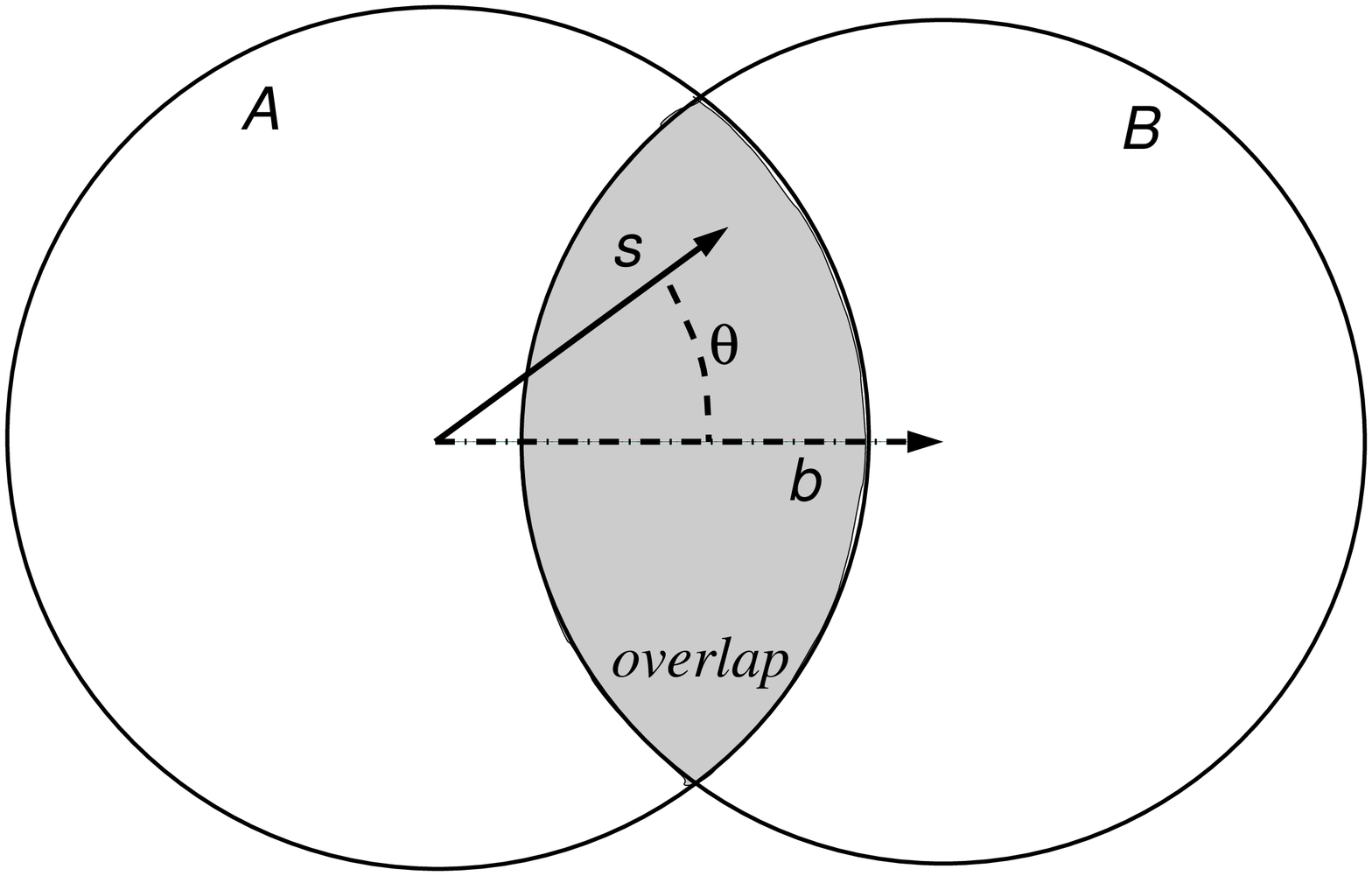,scale=0.26}\hspace{0.7cm}
   \epsfig{file=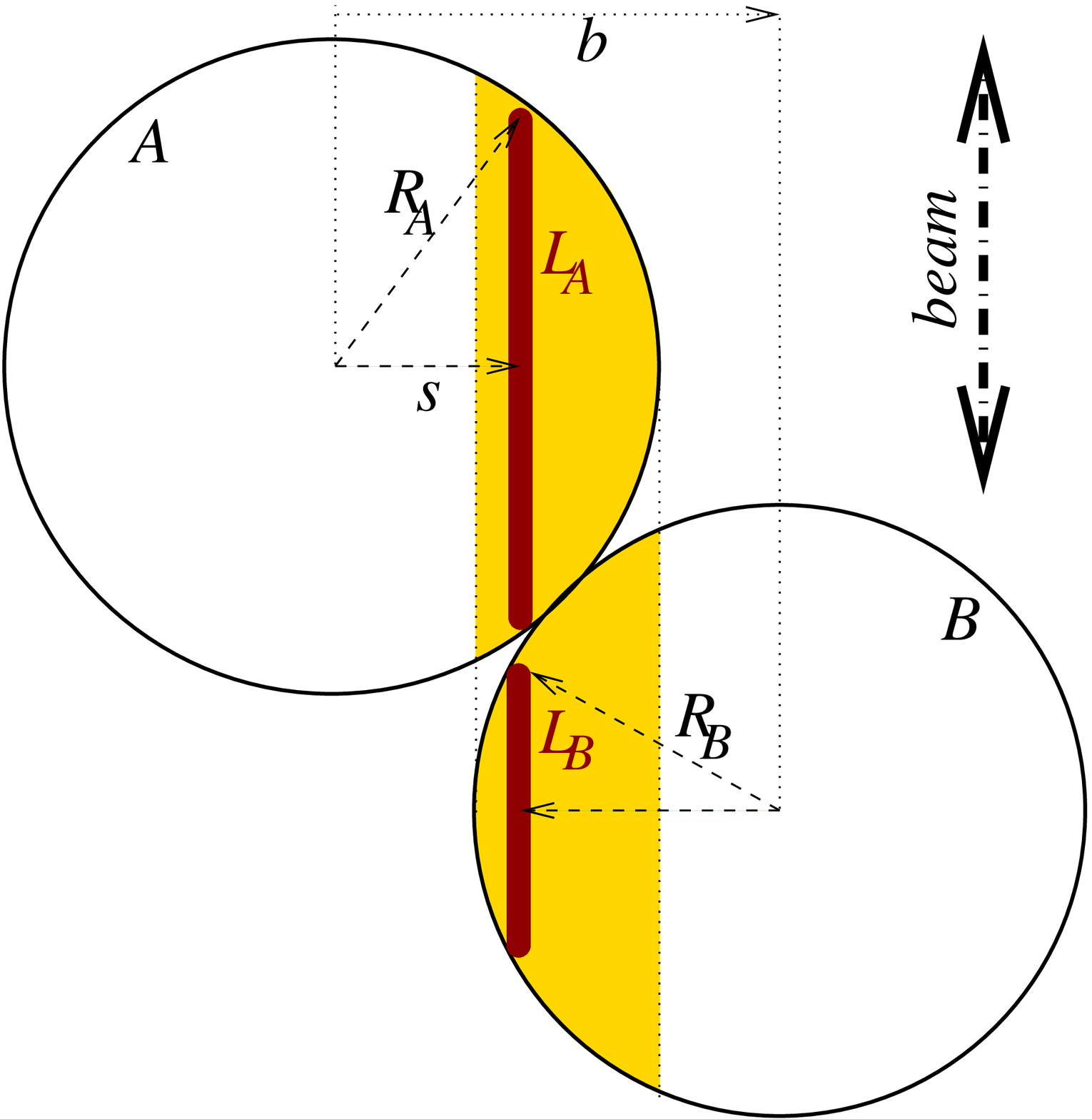,scale=0.3}
\end{center}
\caption{%
Left: geometry of non-central nuclear collisions. Right: 
layout of tube-on-tube interaction (plotted without Lorentz 
contraction).}
\label{f-geom}
\end{figure}
For the sake of simplicity we shall take the nuclei as hard spheres
with radii $R_A=1.2A^{1/3}$ fm and homogeneous number density 
$\rho = 0.138\,\mbox{fm}^{-3}$. In the transverse plane, the impact 
parameter is denoted as $\vec b$ and a point in the transverse plane 
of the nucleus A is specified by the transverse coordinate $\vec s$.
The angle between $\vec b$ and $\vec s$ is denoted as $\theta$.
The situation is sketched in Fig.~\ref{f-geom}.
In a Glauber model, the first part of the nuclear collision is
described as a sum of tube-on-tube interactions (Fig.~\ref{f-geom}). 
In our model we will be interested in the energy density contained 
in such tubes just after the interaction.

The cross-section of both tubes is equal to the non-diffractive 
nucleon--nucleon cross-section $\sigma = 30\,\mbox{mb} = 3\,\mbox{fm}^2$. 
The lengths of the tubes $2L_A$ and $2L_B$ are given by
\begin{eqnarray}
2L_A(s) & = & 2\sqrt{R_A^2-s^2}\, ,\\ 
2L_B(b,s,\theta) & = & 2\sqrt{R_B^2-b^2-s^2+2bs\cos\theta}\, .
\label{eq1}
\end{eqnarray}
The average numbers of nucleons in both tubes are
\begin{equation}
\langle n_A(s) \rangle = 2L_A(s)\rho\sigma,\quad 
\langle n_B(b,s,\theta) \rangle = 2L_B(b,s,\theta)\rho\sigma\, .
\label{eq2}
\end{equation}

For this type of models three ingredients  have to be specified:
\begin{itemize}
\item[(i)] 
The probability distribution of the number of nucleons within the 
colliding tubes $P_A(n_A)$ and $P_B(n_B)$.
\item[(ii)]
For given $n_A$, $n_B$ the nucleons in the tube in A can be 
numbered, starting with the head of the tube as 
$i = 1,\, 2,\, \dots,\, n_A$,
and similarly in B $j = 1,\, 2,\, \dots,\, n_B$. It is
assumed that every nucleon in the tube in A collides with
every nucleon in the tube in B. We have to specify the 
rapidity distribution of all nucleons before and after
every nucleon--nucleon collision.
\item[(iii)]
Finally, we have to specify the production of secondary 
particles and  compute their energy distribution in rapidity after
every nucleon--nucleon collisions. Consider the collision of
the $i$-th nucleon in the tube in A, with the $j$-th nucleon in the 
tube in B, which we refer to as an $(i,j)$ collision. 
Both nucleons have lost a part of their rapidity in interactions
prior to the $(i,j)$ collision.
If we denote the incoming rapidities in such a collision by
$y^i_A$ and $y^j_B$, we need
to specify $dN(y^i_A,y^j_B)/dy\, dp_T^2$ or at least the 
$p_T$-integrated distribution $dN(y^i_A,y^j_B)/dy$ of secondaries
produced in the $(i,j)$ collision. From the rapidity spectrum one 
computes the rapidity distribution of energy 
$\Delta E^{(ij)}/\Delta y$ of the produced particles within the 
interval $-0.5 < y < 0.5$.
\end{itemize}

After having specified the items (i)--(iii), the energy
contained in all {\em secondary} particles produced from a collision 
of two {\em tubes} in the rapidity interval 
$-0.5 < y < 0.5$ is obtained as 
\begin{equation}
\frac{\Delta E^{sec}}{\Delta y}=\sum_{n_A,n_B = 0}^{\infty} P_A(n_A)
\, P_B(n_B) \sum_{i=1}^{n_A}
\sum_{j=1}^{n_B}\frac{\Delta E^{(ij)}}{\Delta y}\, .
\label{eq3}
\end{equation}
In order to obtain the total energy within the given rapidity 
interval we have to add the energy of incident nucleons
in the two colliding tubes, which end up 
in the rapidity interval  $-0.5 < y < 0.5$ when the tube-on-tube collision
is finished. We denote this
contribution by the index ``st'' from nucleon stopping and obtain
\begin{equation}
\frac{\Delta E^{tot}}{\Delta y}= \frac{\Delta E^{sec}}{\Delta y}
+ \frac{\Delta E^{st}}{\Delta y} \, .
\label{eq4}
\end{equation}
This is the total energy resulting from a tube-on-tube collision.
We will have to divide it by the volume occupied by quanta which 
were produced from these two tubes.

We shall now describe the three inputs (i)--(iii) in our model. 

\paragraph{(i) Number distribution.} 
The distribution of the number of nucleons in both tubes
is assumed to be  Poissonian,  with mean values
$\mu_A=\langle n_A(s)\rangle=2L_A(s)\rho\sigma$ and similarly for
$\mu_B$ 
\begin{equation}
P_A(n_A)=\frac{(\mu_A)^{n_A}\exp(-\mu_A)}{n_A!}, \quad
P_B(n_B)=\frac{(\mu_B)^{n_B}\exp(-\mu_B)}{n_B!} \, .
\label{eq5}
\end{equation}

\paragraph{(ii) Rapidity loss.}
In each nucleon--nucleon collision the rapidity loss 
of both nucleons is $\Delta y$. We will show results calculated
for $\Delta y = 0.5$ and $\Delta y = 0.7$. In the CMS of nucleon--nucleon
collisions at the CERN SPS, the absolute value of the rapidity of 
incident nucleons is
$y = | y_A^0 |  = | y_B^0| \approx 3$. In every collision it decreases
by $\Delta y$.  When the rapidity of a nucleon 
becomes less than $\Delta y$, the nucleon does not participate in further
collisions and its energy contributes to $\Delta E^{st}$
in Eq.~\eqref{eq4}.
The term $\Delta E^{st}/\Delta y$ is thus estimated as
\begin{equation}
\frac{\Delta E^{st}}{\Delta y}=\sum_{n_A,n_B=0}^{\infty} 
P_A(n_A)\, P_B(n_B)\, 
\left [ n_{A,slow}(n_A,n_B)+n_{B,slow}(n_A,n_B)\right ]\times 
1\, \mbox{GeV}\, ,
\label{eq13}
\end{equation}
where $n_{A,slow}$ and $n_{B,slow}$ are numbers of incident nucleons 
which end up in the final state with $- 0.5 < y < 0.5$.

\paragraph{(iii) Particle production.}
We assume that the production of secondaries is the same 
as it is in vacuum and 
use the parametrization due to Wong and Lu \cite{Wong3} for the
computation of the energy of secondary particles produced 
in the $(i,j)$ collision. In this model, the rapidity of 
charged particles produced in a collision of nucleons
with rapidities $y_A^i$ and $y_B^j$ is given as
\begin{equation}
\frac{dn^{ij}}{dy}=A\left( (1-x_+)(1-x_-)\right)^a \, ,
\label{eq6}
\end{equation}
where
\begin{subequations}
\label{pp-param}
\begin{equation}
x_+=\frac{m_{\pi T}}{M_N}\exp(y-y^i_A), \quad
x_-=\frac{m_{\pi T}}{M_N}\exp(y^j_B-y),
\label{eq7}
\end{equation}
\begin{eqnarray}
a& = & 3.5+0.7\ln\sqrt{s_{ij}} \, ,
\label{eq9}\\
m_{\pi T}& = & (m_{\pi}^2+B_T^2)^{1/2} \, ,
\label{eq10}\\
A & = & 0.75+0.38\ln\sqrt{s_{ij}} \, ,
\label{eq11}\\
B_T & = & 0.27+0.037\ln\sqrt{s_{ij}} \, .
\label{eq11a}
\end{eqnarray}
\end{subequations}
In Eqs.~\eqref{pp-param}, $m_{\pi}$ is the pion mass and $M_N$ 
the nucleon mass. The average transverse momentum
of produced pions $B_T$ is taken in units of GeV/$c$ and 
the CMS energy of the $(i,j)$ collision $\sqrt{s_{ij}}$ in 
units of GeV.  As discussed in \cite{Wong3}, 
the parameters in \eqref{pp-param} were tuned
by comparison with experimental data \cite{R38,R40}.

The energy of neutral secondary particles (mostly pions) in 
the final state is taken into account by assuming that 
$n_{\pi^0} = (n_{\pi^+} + n_{\pi^-})/2$ and multiplying the factor $A$
in Eq.~\eqref{eq11} by 3/2. In order to go from particle distribution
to energy distribution we multiply the right-hand side 
of Eq.~\eqref{eq6} by the mean pion energy 
$\langle E_{\pi}\rangle$ and obtain
\begin{equation}
\frac{\Delta E^{ij}}{\Delta y}=
1.5A \left((1-x_+)(1-x_-)\right)^a \langle E_{\pi}\rangle\, .
\label{eq12}
\end{equation}
Here
$\langle E_{\pi}\rangle = (m_{\pi}^2 + B_T^2 +p_L^2)^{1/2}$,
with $p_L^2 = \langle p_T^2 \rangle /2 = B_T^2/2$ for pions 
in the rapidity interval $-0.5 < y < 0.5$.
The value of $\Delta E^{ij}$/$\Delta y$ calculated by Eq.~\eqref{eq12}
is then inserted into  Eqs.~\eqref{eq3} and \eqref{eq4}.

Our interest here is in the energy density of 
quanta which form the system when the tube-on-tube interactions 
are finished. These quanta include secondary particles produced 
according to Eq.~\eqref{eq6} and slowed-down nucleons. The 
energy density is given by
\begin{equation}
\eden = \frac{\Delta E^{tot}/\Delta y}{V(L_A,L_B,\Delta y)} \, ,
\label{eq14}
\end{equation}
where  $V(L_A,L_B,\Delta y)$ is the volume occupied by the central rapidity 
unit when the collision of the two tubes is finished. It is only the
volume of particles involved in a single tube-on-tube process, not 
the total fireball volume. We estimate it as
\begin{equation}
V(L_A,L_B,\Delta y)= (2L_A/\gamma+2L_B/\gamma+2v_0t_0)\sigma\, .
\label{eq15}
\end{equation}
Here, $\gamma$ is the Lorentz contraction factor, 
for Pb+Pb collisions at the CERN SPS
$\gamma \approx 9$.  The third term $2 v_0 t_0$ stands for the delay 
due to formation time. We assume 
that the formation of particles effectively sets in after
the two colliding tubes have crossed each other.
In our calculation
we put $v_0 = 0.5$ and varied the parameter $t_0$.




\section{Comments on the model}
\label{comments}

The Glauber model of pA and AB interactions has been studied by many 
authors 
\cite{Date,Wong1,Wong2,Wong3,Sumi1,Sumi2,Lexus,Satz1,Liet1,Liet2,Zav,Lich,Gale,Hwa1,kapust,huf}.
Experimental data have been reviewed by Busza and Ledoux \cite{Busza}. 
As pointed out in \cite{Busza},
one of the problems of this field is caused by insufficient 
accuracy of the data. The situation can change soon, since the NA49 
Collaboration at the CERN SPS has recently obtained new data
on pA interactions, which were only briefly published so far 
\cite{Blume,Fischer:2002qp}. 
Note that these data are in the same energy region where 
anomalous $J/\psi$ suppression \cite{NA50a} occurs and 
increased production
of multistrange baryons \cite{NA57a} has been seen. 
For a more precise formulation of the model 
one would need accurate data on nucleon stopping in pA interactions,
on the production of secondary particles in multiple collisions of
a proton in the nucleus, and on the formation time of secondary
particles. 

We shall now discuss the assumptions made in our simple model as well
as the choice of parameters. 

First we turn to our assumption about the rapidity loss
in every nucleon--nucleon collision.
In pp collisions, the proton rapidity loss is as large as 
$\Delta y \approx 1$ \cite{R38,Busza} and this is assumed
in many models \cite{Lexus,Satz1,Zav,Wong3,Hwa1,huf} in which the 
data on pp 
interactions are directly extended to multiple collisions in pA 
and AB interactions. In \cite{Liet1,Liet2} 
the rapidity loss is put equal to $\Delta y \approx 0.5$ and in some 
other models \cite{Date} the 
rapidity loss per collision is as small as $\Delta y \approx 0.3$.
Recent data of the NA49 Collaboration \cite{Fischer:2002qp}
support our choice of $\Delta y = 0.5$. In order to see the influence 
of a larger $\Delta y$, we have also made calculations 
with $\Delta y = 0.7$
which is closer to the assumptions made in most models.

A crucial assumption made in our model is the way in which the volume 
$V(L_A,L_B,\Delta y)$ is determined, see Eq.~\eqref{eq15}. 
We discuss this point in more detail here.

An important contribution to the volume in Eq.~\eqref{eq15} comes from the
formation time. We want to note that in a pA interaction the collision of the 
incident proton with the tube in the nucleus is a very complicated 
process which we are unable to describe in detail. A description of this
process by multiple collisions only gives the final
state but does not make statements about the real 
intermediate stages of the process itself. 
In early critical comments on this type of description \cite{Jar} it 
was already pointed out that a simple classical explanation 
contradicts the data on Drell--Yan production in pA and AB
interactions. The argument goes along these lines: suppose that in the first 
collision in the nucleus the proton loses some part of its 
momentum and the parton structure functions immediately 
adapt to this change. In such a situation the cross-section 
for the production 
of Drell--Yan pairs cannot be proportional to A$^{\alpha}$ with 
$\alpha$ very close to 1. The same problem appears in the case of 
Drell--Yan production in nuclear collisions, where the cross-section is
again proportional to (AB)$^{\alpha}$ with $\alpha$ close to 1.
This issue has been recently discussed by Gale, Jeon and Kapusta \cite{Gale}.
They have introduced a coherence time of the energy loss of the proton in pA
interactions. This coherence time gives the delay after which
the proton energy is degraded to the value seen in the final state.
By analysing the data \cite{Alde} on Drell--Yan
pair production in pA interactions, they concluded that 
the average proper coherence time $t_0 = 0.4 \pm  0.1\, \mbox{fm}/c$.
They have also pointed out that this coherence time is related to the 
formation time of secondary hadrons in pA interactions. Indeed, when a 
secondary final-state hadron is able to interact with other particles,
its energy must be already felt as lost by the incident proton.
This estimate of the formation time may rather be 
considered as a lower limit, since a part of the data in 
\cite{Alde} can
be explained by shadowing corrections and by the energy 
loss of incident partons when traversing the nucleus \cite{Kope}.
Possible effects of shadowing and energy loss corrections
in the Drell-Yan data from Fermilab were discussed for the 
first time in \cite{STDY}. Recent analyses can be found in
\cite{Kope}.

Although the model of pA interactions uses hadronic
degrees of freedom, this does not imply their dominance
in the intermediate stages of the process.
Since the model refers only to the final state, it is quite possible 
that the dynamics of the 
intermediate stages is dominated by partonic degrees of freedom.
The relationship between intermediate and final stage is
thus given by some form of parton--hadron  duality.

Let us come back to our estimate of the volume.
In our scenario, the process of tube-on-tube collision can be 
finished only after 
both tubes traverse each other completely. The effective volume is
further increased by the delay needed for the formation of
secondaries, which move with rapidities up to $y=\pm 0.5$.
The value of the formation time $t_0$ is assumed to be of
the order of 1 fm/$c$ and is a free parameter. 
This estimate
is based on works quoted in \cite{PPZ}, in particular on 
\cite{Ranft}. For experimental data, see \cite{Brick}.
The data underlying these estimates are rather old and 
a new and more accurate information on the value of the formation time
is most desirable.

Before finishing this section, let us discuss some alternative 
estimates of the volume.
One could argue that most of the secondaries are produced in 
the early nucleon--nucleon collisions and therefore the total lengths 
of the tubes $2L_A$ and $2L_B$ should be replaced by some 
effective lengths, say $l^{\rm eff}_A \approx L_A$
and $l^{\rm eff}_B \approx L_B$. Then
\begin{equation}
V^{\rm half}(L_A,L_B,\Delta y) = (L_A/\gamma+L_B/\gamma+2v_0t_0)\sigma\, .
\label{eq16}
\end{equation}
On the other hand, the process is
not finished when only the front halves of the tubes cross each
other, and the production of secondary particles is
disturbed by nucleons that come later.

A very popular Bjorken scenario \cite{Bj} has been devised
for asymptotic energies. In this regime the nuclei are really contracted
to pancakes of vanishing thickness.
%
%
In the CMS of Pb+Pb interactions at the SPS the
nuclei are contracted to $2R_{\rm Pb}/\gamma \approx 1.58\, \mbox{fm}$ 
and neglecting this length is not realistic.

Both these estimates would lead to smaller volumes than 
formula \eqref{eq15} and hence to larger energy densities.
In what follows, however, we only report on results obtained 
with the estimate \eqref{eq15}, which we consider as most realistic.


\section{Results}
\label{results}

\begin{table}
\caption{Largest length of tubes for some selected nuclei in units of
fm.
\label{tab1}} 
\begin{center}
\begin{tabular}{|c|ccccccccc|}\hline
A&$^{12}$C&$^{16}$O&$^{32}$S&$^{64}$Cu&$^{115}$In&$^{108}$Ag&$^{184}$W 
& $^{208}$Pb & $^{238}$U \\ 
$2L_A^{max}$ &5.49&6.05&7.62&9.6&11.67&11.43&13.65&14.22&14.87 \\ \hline
\end{tabular}
\end{center}
\end{table}
We calculated the energy density as a function of the length of the 
tubes. Before plotting our results it is useful to quote
realistic values for the maximum length of tubes that fit into 
a few selected nuclei. These lengths are given by 
$2L_A^{max} = 2 R_A = 2.4 A^{1/3}$ and are summarized in 
Table~\ref{tab1}.

\begin{figure}[t]
\begin{center}
   \epsfig{file=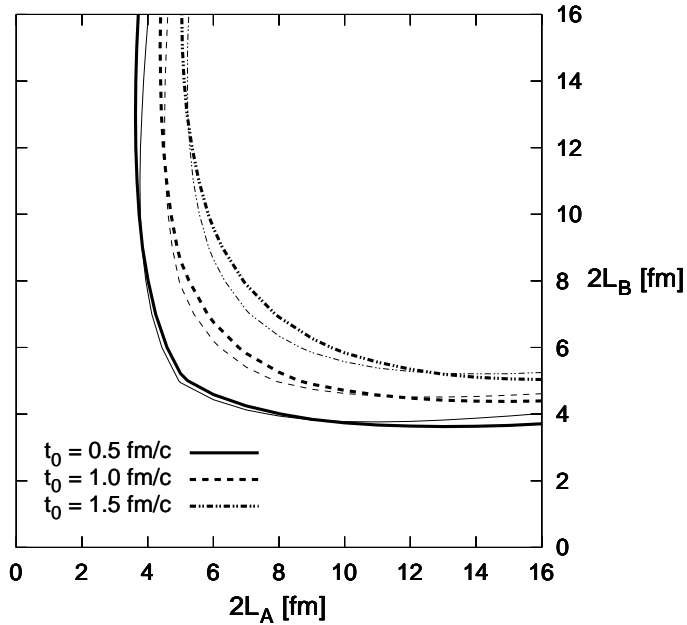,scale=0.65}
\end {center}
\caption{
Lines of constant energy density 
$\eden=\eden_c= 0.7\,\mbox{GeV/fm}^{3}$ calculated 
for $t_0 = 0.5, \, 1.0,\, 1.5\, \mbox{fm}/c$. 
Nucleon stopping was set to $\Delta y= 0.5$ (thick lines) and
$\Delta y=0.7$ (thin lines). Other
parameters are given in the text. Energy density is calculated by
Eqs.~\eqref{eq14} and \eqref{eq15}.  
}
\label{fig1}
\end{figure}

In all our calculations the following choice of parameters is made:
$\sigma= 30\, \mbox{mb}$, $\gamma=9$, $\rho= 0.137\,\mbox{fm}^{-3}$, 
$y_A^0=3$, $y_B^0=-3$.

In Fig.~\ref{fig1} we plot six curves corresponding to the critical 
energy density $\eden=\eden_c= 0.7\,\mbox{GeV/fm}^{-3}$ as a function 
of $2L_A,\, 2L_B$. They were obtained for $\Delta y = 0.5$ and 0.7, and 
different values of the formation time parameter $t_0$. 

The influence of stronger nucleon stopping on the results is very weak:
there is only little modification in the curves as seen from 
Fig.~\ref{fig1}. This demonstrates that the major contribution
to the energy density comes from produced particles, i.e.\ the 
first term in Eq.~\eqref{eq4}. Quite naturally, longer formation
times lead to larger volumes and lower energy densities, and
the critical energy density is thus reached in collisions of longer tubes.

The results are very optimistic. According to  
Table~\ref{tab1}, critical energy density is just reached in the centre
of head-on S+S collisions and certainly in S+Pb or S+U interactions,
because the combination of longest tube lengths for these pairs
of nuclei falls into the region above the critical curve in 
Fig.~\ref{fig1}. One is tempted to claim the
existence of the QGP. However, if we accept the conjecture that 
anomalous $J/\psi$ suppression is connected with plasma production,
our statement comes out too optimistic. Anomalous $J/\psi$ suppression
was only observed for larger collision systems \cite{NA50a}.

$J/\psi$ suppression as a signature of QGP formation in nuclear 
collisions has been proposed by Matsui and Satz \cite{MS86} more 
than 15 years ago. The anomalous $J/\psi$ suppression has been 
discovered by the NA50 collaboration in 1996 \cite{NA50a}. About 
a year later phenomenological models of the QGP formation 
and $J/\psi$ suppression in nuclear collisions at the SPS have 
been proposed by two groups: 
Blaizot, Ollitrault \cite{BO} (we will refer to this model 
as BO) and 
Kharzeev, Louren\c{c}o, Nardi and Satz (KLNS, \cite{KLNS}). 
As discussed in Refs.~\cite{BO,KLNS} and by Nogov\'a {\it et al.}
\cite{NPP00}, the condition for QGP formation in the BO model 
can be stated simply as
\begin{equation}
\mbox{[BO]}\qquad \qquad 2L_A(s) + 2L_B(b,s,\theta) \ge 23.5\,\mbox{fm}\, .
\label{eq18}
\end{equation}
For the KLNS model the condition for the formation of QGP reads
\begin{equation}
\mbox{[KLNS]} \qquad \qquad
2L_A \times 2L_B \ge 5.87\, \mbox{fm} \times (2L_A+2L_B) \, .
\label{eq19}
\end{equation}
In these equations, $2L_A,\, 2L_B$ are given in units of fm. 
\begin{figure}[t]
\begin{center}
   \epsfig{file=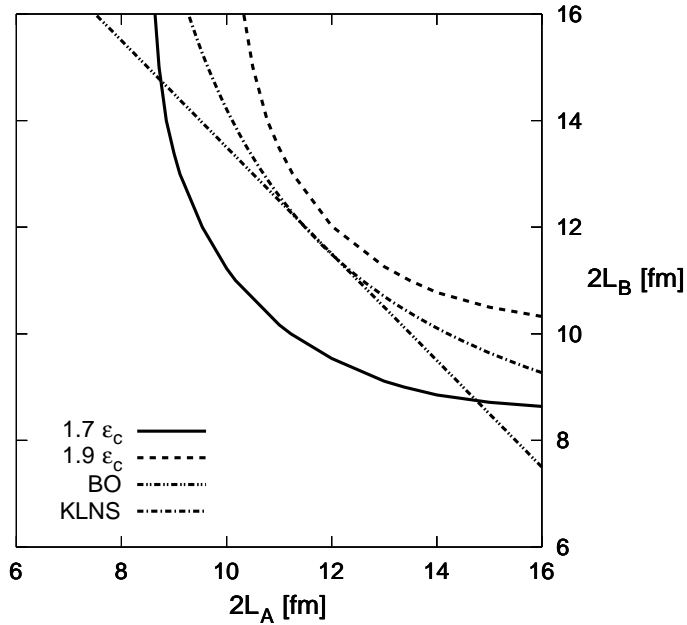,scale=0.65}
\end {center}
\caption{%
Curves in the 2$L_A$, $2L_B$ plane, which indicate the 
limits of anomalous $J/\psi$ suppression in the BO \protect\cite{BO}
and KLNS \protect\cite{KLNS} models, given by
Eqs.~\eqref{eq18} and \eqref{eq19}, respectively. 
These curves are compared 
with constant energy density curves with $\eden = 1.7 \eden_c$
and $\eden = 1.9 \eden_c$, which were calculated in our model
for $\Delta y = 0.5$ and $t_0 = 1\, \mbox{fm}/c$.}
\label{fig3}
\end{figure}
The critical curves for plasma formation according to these two 
models are shown in Fig.~\ref{fig3}. Note that for interactions
of equal-length tubes they both lead to almost the same results. 
Differences
can only show up in non-central collisions and interactions of 
different size nuclei.

We compare with these curves the results of our energy density 
estimate with reasonable parameters values. The critical curves
of BO and KLNS models correspond to an energy density about 
1.8 times above $\eden_c$. We interpret this result as a consequence
of delayed thermalization and  rapid expansion in both the 
longitudinal and  transverse \cite{HT1} direction
which leads to fast cooling. In order to have a collectively behaving
plasma that will be 
able to screen the $c\bar c$ interaction, particle production 
must lead to an energy density  higher than $\eden_c$.
Only in such a case can it remain in a deconfined state until the
collective behaviour is established.


\section{Comments and conclusions}
\label{conclusions}

Our estimates of energy densities in heavy-ion collisions in the CERN
SPS energy region have been based on the Glauber-type model 
of tube-on-tube interactions. The tube-on-tube prescription 
of nuclear collisions was complemented by the introduction 
of the formation time.

In spite of our rather conservative choice of model parameters,
the energy densities we obtained are higher than the critical 
value known from lattice gauge theory.  
Thus our results support the claim 
that the threshold for quark--gluon plasma formation has been reached at  
the CERN SPS \cite{HJ}. This statement requires a few comments.
\begin{itemize}
\item[(i)]
The lattice results \cite{Karsch1,Karsch2} on the critical energy 
density still  have rather large error bars. According to \cite{Karsch1} 
$\eden_c = (6 \pm 2) \times (173 \pm 8)\, \mbox{MeV}^4$ which
leads to $\eden_c  = 0.7 \pm  0.35 \, \mbox{GeV/fm}^3$. In addition to
statistical errors on $T_c$, there are also systematic errors of a 
comparable size due to extrapolation to the chiral limit.
The uncertainty due to statistical errors is illustrated in 
Fig.~\ref{f:band} where we plot 
the band between the curves for $\eden_c = 0.35 \, \mbox{GeV/fm}^3$ 
and $\eden = 1.05\, \mbox{GeV/fm}^3$ corresponding to
our results for $\Delta y = 0.5$ and $t_0 = 1\,\mbox{fm}/c$, together with
the curves obtained in the BO and KLNS models.
The curve corresponding to
$\eden_c = 1.05\, \mbox{GeV/fm}^3$ (that means 1.5 
times 0.7 GeV/fm$^3$) starts to approach the BO and KLNS models.
In view of this uncertainty it becomes unclear whether or not critical
energy density is reached in S+S collisions, but central collisions
of In+In and heavier systems appear to be on the safe side.
%
\begin{figure}[t]
\begin{center}
   \epsfig{file=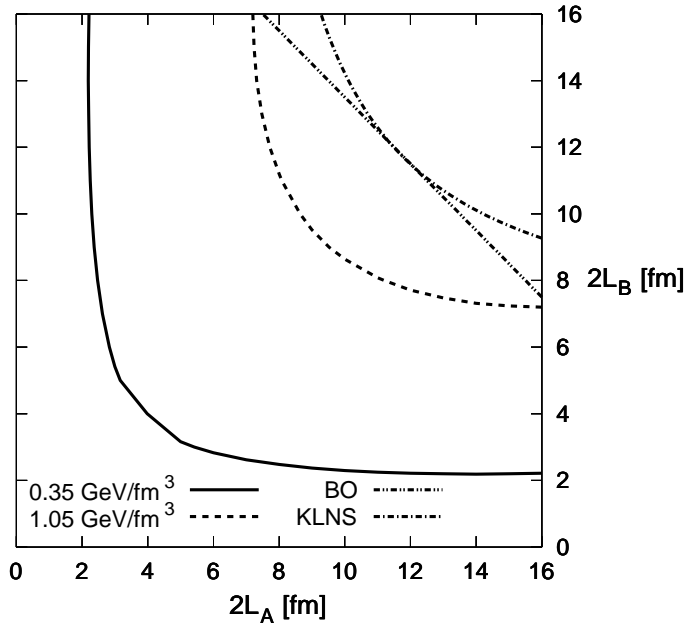,scale=0.65}
\end {center}
\caption{%
The band in $2L_A$, $2L_B$ plane corresponding to
$\eden_c = 0.7 \pm 0.35\,\mbox{GeV/fm}^3$, calculated for
$t_0 = 1\, \mbox{fm}/c$ and $\Delta y = 0.5$.
}
\label{f:band}
\end{figure}
%
\item[(ii)]
We have tacitly assumed that the local energy density as 
calculated above for a small size system can be compared with 
lattice results corresponding to an infinite system.
\item[(iii)]
The model we are using is based on a phenomenological picture of
production of (mostly) soft hadrons. Although the model has been 
tested by comparison with the data on hadron production, the true
dynamics of the process might be somewhat different. It is for 
instance possible \cite{ABU}, that the first stage of the collision 
is dominated by production of gluons 
with momenta of about 0.6--1.0 GeV/$c$ in nucleon-nucleon
interactions and the system---depending on the energy 
density---either hadronizes or approaches kinetic equilibrium. In the latter
case the formation time as used in Eqs. (11) and (12) should rather 
correspond to the thermalization time. This can be roughly estimated
as the time (in the c.m.s. of colliding nucleons) required for the
emission of a few softer gluons by the harder ones originally 
produced. Since the emission of a gluon with momentum of $kT_c$
takes about 1 fm/$c$, the approach to equilibrium may take 2--3 fm/$c$.
Larger formation time would shift our results in 
Fig.~\ref{fig1} closer
to the curves obtained in BO and KLNS models from phenomenological 
analyses of data on anomalous $J/\psi$ suppression. 
This is seen 
in Fig.~\ref{fig6} where we present the energy density obtained in our model
for values of formation times of 2 and 3 fm/$c$. 
Also note  that
lattice results correspond to the system in equilibrium and 
very little 
is known about  $J/\psi$ suppression by partonic systems out of
equilibrium.
\begin{figure}[t]
\begin{center}
   \epsfig{file=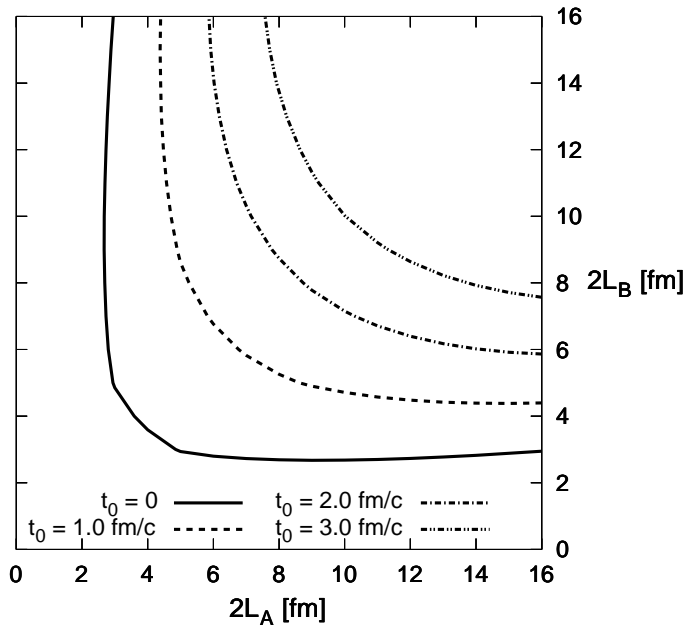,scale=0.65}
\end {center}
\caption{%
Curves of the critical energy density calculated 
for formation (equilibration) times $t_0 = 0,\, 1,\, 2,\, 3\, \mbox{fm}/c$, 
and $\Delta y = 0.5$.
}
\label{fig6}
\end{figure}
%
\item[(iv)]
The model used is based on the assumption of purely 
longitudinal dynamics of the nuclear collision---at least in that 
part of the collision which leads to formed hadrons or the system
of gluons close to kinetic equilibrium. The data on HBT radii and
on the transverse momentum spectra in heavy-ion collisions at the
CERN SPS indicate the presence of a rapid onset of the transverse
flow \cite{HT1,PrattQM}. Transverse expansion lowers the energy density
(with respect to the one calculated in a purely longitudinal
dynamics) and this would move our curves closer to those obtained
in the BO and KLNS models. Unfortunately, without a more detailed
information about the time evolution of the
transverse flow it is difficult to estimate the effect.
\item[(v)]
It has been shown by the Bielefeld group \cite{DIGAL} and later
confirmed by Wong and co-workers 
\cite{WONGG} that (2S) and (1P) quarkonia may
already be dissolved below $T_c$. This affects the observed $J/\psi$
suppression via dissolution of (1P) quarkonia since about 40\% of 
$J/\psi$'s in the final state is due to radiative decays of $\chi$'s.
Hence, anomalous $J/\psi$ suppression starts at energy density lower
than $\eden_c$ The discrepancy between the lines corresponding to 
this lower energy density in our model 
and the BO and/or KLNS models shown in 
Fig.~\ref{fig3} is yet higher than what is presented in the Figure.
\end{itemize}

To summarize: we have computed energy densities reached in heavy-ion 
collisions at the CERN SPS in a simple model based on the
assumption of longitudinal dynamics in the stage preceding 
hadronization or kinetic equilibration. For formation times 
of about 1~fm/$c$ the energy density of $\eden_c = 0.7\,\mbox{GeV/fm}^3$
is reached in collisions of tubes which are shorter than 
assumed in phenomenological models of Blaizot and Ollitrault
and of Kharzeev, Louren\c{c}o, Nardi and Satz. The discrepancy
is most likely due to combination of the three following
effects: larger formation or equilibration times than usually
assumed, true critical energy density larger than 
0.7 GeV/fm$^3$
and a possible rapid onset of transverse expansion.

\paragraph{Acknowledgements}
One of the authors (JP) is
indebted to the CERN Theory Division for the hospitality 
extended to him. BT is grateful to P.~Vagner and M.~Mo\v sko 
of the Slovak Academy of Sciences where a part of this paper
was written for their hospitality.
We would like to thank F.~Karsch, C.~Salgado and
U.~Wiedemann for comments and helpful discussions. 
We are also indebted to the unknown referee for remarks 
and recommendations which have improved and clarified
the presentation of this paper.
The work of NP 
and JP has been supported in part by the grant of the Slovak Ministry 
of Education No.\ VEGA V2F13.

\end{document}